| LETTER |
| --- |

# The Impact of Defect (Re) Prediction on Software Testing

Yukasa MURAKAMI†, Yuta YAMASAKI††, *Nonmembers*, Masateru TSUNODA††,
Akito MONDEN†, *Members*, Amjed TAHIR†††, Kwabena Ebo BENNIN††††, *Nonmembers*,
Koji TODA†††††, *Member*, and Keitaro NAKASAI††††††, *Nonmember*

**SUMMARY** Cross-project defect prediction (CPDP) aims to use data from external projects as historical data may not be available from the same project. In CPDP, deciding on a particular historical project to build a training model can be difficult. To help with this decision, a Bandit Algorithm (BA) based approach has been proposed in prior research to select the most suitable learning project. However, this BA method could lead to the selection of unsuitable data during the early iteration of BA (i.e., early stage of software testing). Selecting an unsuitable model can reduce the prediction accuracy, leading to potential defect overlooking. This study aims to improve the BA method to reduce defects overlooking, especially during the early testing stages. Once all modules have been tested, modules tested in the early stage are re-predicted, and some modules are retested based on the re-prediction. To assess the impact of re-prediction and retesting, we applied five kinds of BA methods, using 8, 16, and 32 OSS projects as learning data. The results show that the newly proposed approach steadily reduced the probability of defect overlooking without degradation of prediction accuracy.

## 1. Introduction

Software testing is a critical step in discovering and removing defects. However, testing can be less frequent due to the limited resources (especially human effort and time) [12]. Defect prediction models are applied to find potential defects easily and early in the testing phase. When a module is regarded as defective by the prediction model, testing resources can be allocated to such modules for thorough testing [9]. Thus, improving the accuracy of prediction models can lower testing efforts and improve software quality.

Data collected on the previous version of the prediction target software is often used to build a defect prediction model. However, newly built software will not have any training data for the prediction model. A feasible solution is to use data collected from other software projects (obtained internally or externally). This is referred to as cross-project defect prediction (CPDP). CPDP has attracted increased attention in recent years [4]. However, the characteristics of software projects can vary from one project to another. CPDP models trained on arbitrarily selected projects different from the target project do not perform well [11].

Still, there are challenges in identifying suitable projects for data training [8]. To help with the selection, a Bandit Algorithm (BA) based approach has been proposed to select the most suitable learning project [2]. However, this BA method could lead to the selection of unsuitable data during the early iteration of BA (i.e., early stage of software testing). Selecting an unsuitable model can reduce the prediction accuracy, leading to potential defect overlooking. The study aims to improve the BA method to reduce defects overlooking, especially during the early stage of testing.

## 2. Bandit Algorithm (BA) Based Defect Prediction

**Overview**: Our previous work has extensively discussed bandit algorithm (BA) based defect prediction [10]. The BA method assumes the following:

- **B1**: Each module is tested sequentially during testing.
- **B2**: The test result of each module is recorded.

Except for "big-bang" Integration testing, each module is tested sequentially during the testing phase [1], and results are recorded - even when we do not apply the BA method. Therefore, most software development satisfies B1 and B2.

The BA-based method builds prediction models using data from different projects as learning data. During software testing, the model is not rebuilt. Therefore, selecting one of the models means selecting the learning data. In Fig.1, four prediction models are built before testing, using data collected from projects A, B, C, and D as learning data. In the figure, 100 modules are sequentially tested, and the numbers in parenthesis signify the test order of the modules. In this case, module t21 is the test target module, and gray rows signify tested modules. ND, DE, CO, and WR mean non-defective, defective, correct, and wrong, respectively.

As shown in Fig. 1, the BA method selects a higher-accuracy model by performing the following procedure.

Step 1.   Select a model randomly.
Step 2.   Use the prediction of the selected model.
Step 3.   Test the module and record the result.

———————————————————
† The authors are from Okayama University, Japan.
†† The authors are with Kindai University, Japan.
††† The author is with Massey University, New Zealand.
†††† The author is with Wageningen University & Research, the Netherlands.
††††† The author is with Fukuoka Institute of Technology, Japan.
†††††† The author is with Osaka Metropolitan University College of Technology, Japan.

| Test Module (order) | Prediction | | | | Selected model (Prediction) | Test result | Evaluation on test | | | | | | | |
|---|---|---|---|---|---|---|---|---|---|---|---|---|---|---|
| | Model A | Model B | Model C | Model D | | | Model A | Model B | Model C | Model D | AUC of A | AUC of B | AUC of C | AUC of D |
| ... | ... | ... | ... | ... | ... | ... | ... | ... | ... | ... | ... | ... | ... | ... |
| t38 (4) | DE | ND | DE | DE | A (DE) | ND | WR | CO | WR | WR | 0.75 | 0.74 | 0.73 | 0.72 |
| t21 (5) | DE | DE | DE | DE | A (DE) | DE | CO | CO | CO | CO | 0.76 | 0.75 | 0.74 | 0.73 |
| t75 (6) | ND | DE | DE | DE | A (ND) | ND | CO | WR | WR | WR | 0.77 | 0.74 | 0.73 | 0.72 |
| t19 (7) | ND | DE | ND | ND | A (ND) | ND | CO | WR | CO | CO | 0.78 | 0.73 | 0.74 | 0.73 |
| t56 (8) | ND | ND | DE | DE | A (ND) | ND | CO | CO | CO | WR | 0.79 | 0.74 | 0.75 | 0.72 |
| ... | Re-prediction model | | | | | ... | ... | ... | ... | ... | ... | ... | ... | ... |
| t02 (100) | ND | DE | DE | DE | B (DE) | ND | WR | CO | WR | CO | 0.75 | 0.77 | 0.74 | 0.73 |

Fig. 1 Procedure of BA based defect prediction

| Test Module (order) | Selected model (Prediction) | Test result | Test effort | Residual defects | |
|---|---|---|---|---|---|
| ... | ... | ... | ... | ... | |
| t38 (4) | A (DE) | ND | High | No | Case α |
| t21 (5) | A (DE) | ND | High | No | |
| t75 (6) | A (ND) | DE | Low | Yes | Case β |
| t19 (7) | A (ND) | ND | Low | No | |
| t56 (8) | A (ND) | ND | Low | No | |
| ... | ... | ... | ... | ... | |
| t02 (100) | B (DE) | ND | High | Yes | Case γ |

Fig. 2 Relationship between prediction and test result

Step 4. Compare the test result and prediction of each model.

Step 5. Compare the accuracy of each model and select the model with the highest accuracy.

Step 6. Return to Step 2 until all modules are tested.

In step 5, we used AUC to measure prediction accuracy [10]. Several methods can be used to select the models, such as ε-greedy and UCB (Upper Confidence Bound).

**Incorrect selection**: BA's number of comparisons (i.e., accuracy evaluation of predictions) needs to be increased during the early stage of software testing. Therefore, the results could vary when the evaluation increases. For instance, in Fig. 1, model A is selected for module t21, tested fifth. However, model A's accuracy is lower than B when all 100 modules have been tested (e.g., t02 in Fig. 1).

Hence, the prediction during the early stage could be incorrect, as shown in the two cases below:

- **Case α**: When the prediction is defective but the module does not contain defects, the module is tested thoroughly with high effort, but still, no defects are found. As a result, the testing effort increases significantly.

- **Case β**: When the prediction is non-defective, but the module contains defects. The module is then tested lightly with low effort to suppress the total cost of testing [9]. This causes defect overlooking, resulting in residual defects on the module and degrading software quality.

Fig. 2 shows the relationship between prediction and test results. The figure also includes test effort and residual defects, which are mentioned in the explanations of cases α and β. In Fig. 2, test modules and prediction for them are the same as in Fig. 1. In the figure, modules t38 and t75 are case α and β, respectively. After all modules are tested, there is no way to recover the increased effort on case α. In contrast, we could suppress residual defects to some extent in case β by retesting modules thoroughly if we can identify candidates of case β (i.e., defect-overlooked modules).

## 3. Re-prediction and Retesting Approach

**Overview**: We propose re-prediction and retesting to identify case β modules. Our approach assumes the following:

- **R1**: Modules tested earlier could include residual defects due to the lower accuracy of the selected model.

- **R2**: For a module, the cost due to residual defects is higher than retesting the module.

R1 considers that model evaluation is insufficient during the early testing stage. As a result, defect prediction on modules tested during this stage might be inaccurate.

The total effort for a retested module is the sum of testing and retesting efforts for that particular module. Although the former is excessive due to an inaccurate prediction, as shown in Fig. 2, the testing effort is low. This is because fewer test cases are created than cases made for modules that are predicted as defective. When defects are overlooked during a phase but removed in a later phase, the effort of the removal increases excessively [3].

A retest based on defect re-prediction will be performed using the following procedure.

| Test module (order) | Prediction | | | | Re-prediction model (Prediction) | Retest result | Evaluation on retest | | | | | | | | Test effort | Retest effort |
|---|---|---|---|---|---|---|---|---|---|---|---|---|---|---|---|---|
| | Model A | Model B | Model C | Model D | | | Model A | Model B | Model C | Model D | AUC of A | AUC of B | AUC of C | AUC of D | | |
| t38 (4) | DE | ND | DE | DE | B (ND) | - | WR | CO | WR | WR | 0.73 | 0.74 | 0.73 | 0.72 | High | Excessive effort |
| t21 (5) | DE | DE | DE | DE | B (DE) | - | CO | CO | CO | CO | 0.74 | 0.75 | 0.74 | 0.73 | High | |
| t75 (6) | ND | DE | DE | DE | B (DE) | DE | WR | CO | CO | CO | 0.73 | 0.76 | 0.75 | 0.74 | Low | High |
| t19 (7) | ND | DE | ND | ND | B (DE) | DE | CO | WR | CO | CO | 0.74 | 0.75 | 0.76 | 0.75 | Low | High |
| t56 (8) | ND | ND | ND | DE | C (ND) | - | CO | CO | CO | WR | 0.75 | 0.76 | 0.77 | 0.74 | Low | - |
| ... | ... | ... | ... | ... | ... | ... | ... | ... | ... | ... | ... | ... | ... | ... | ... | ... |
| t02 (100) | ND | DE | DE | DE | D (DE) | - | WR | CO | CO | CO | 0.75 | 0.77 | 0.74 | 0.78 | High | - |

**Gray cells**: differences in predictions and their evaluations from previous ones

Fig. 3 Procedure of retesting based on defect re-prediction

Step 1. After all modules have been tested, the re-prediction model is settled based on the accuracy of each model (see Fig. 1).

Step 2. Perform Step 2 of BA if the prediction by the selected model was non-defective (i.e., candidates of case β).

Step 3. If Step 2 is performed, perform Step 3 … 5 of BA when the prediction by the re-prediction model is defective.

Step 4. Return to Step 2 until all modules are re-predicted.

Fig. 3 illustrates this retesting-based procedure. Based on Step 3 (i.e., Step 3 … 5 of BA), the re-prediction model could be changed during this procedure. For instance, the figure changes the model from B to C after retesting module t19.

**Application range**: Note that the application of this proposed approach is not limited to the BA method and CPDP. For instance, we can apply the same concept to CVDP (cross-version defect prediction), which uses data collected during the development of the previous version as learning data. Additionally, as a re-prediction model, we can adopt a new model that uses test results as learning data (i.e., online learning).

**Multiple retests**: After all modules have been re-predicted and retested, we can repeatedly perform the procedure from the first module. For instance, in Fig. 3, if the re-prediction model turns model D on the second iteration of the re-prediction, module t56 is then proposed to be retested because the module was not tested in the first iteration. We call this a multiple retests approach.

## 4. Experiment

**Dataset**: In our experiment, we used data from 33 open-source projects provided in the DefectData dataset[a]. For the test data, we used the arc project. The arc project includes 235 modules, of which 11.5% are defective. We used Chidamber & Kemerer (CK) metrics as candidates for explanatory variables.

As learning data, we randomly selected 8, 16, and 32 pieces of projects from the remaining 32 projects. With many candidates for learning data, it could be difficult for the BA method to select the best learning data. Therefore, we changed the amount of learning data candidates.

**Defect overlooking with "defective" prediction**: Even when the prediction by the model is "defective," some defects could be overlooked. Typically, defects that are discovered after release are considered as overlooked defects. A recent industrial survey [5] reported that about 17% of defects are overlooked during integration testing. The overlooking could occur when the test result is "defective" (and defects might be found during testing and after the software release). We call this **case γ** (see Fig.2). Therefore, similar to [10], to simulate those overlooked defects, we randomly changed the evaluation of BA at 20% probability when the modules are defective.

**Prediction method**: we applied logistic regression to predict defective modules, as it is one of the most widely used methods in CPDP. As a feature selection method, we applied correlation-based feature selection, which is effective when used together with logistic regression [6]. As BAs, we used ε-greedy (ε = 0, 0.1, 0.2, and 0.3) and UCB (Upper Confidence Bound). We compared the CPDP performance of the following approaches:

- **Baseline approach**: Perform the test only with the ordinal BA method
- **Retest approach**: Perform not only the test but also retest with the re-prediction method
- **Multiple retests approach**: Perform the test once and retest twice with the re-prediction method

**Evaluation criteria**: We used AUC to evaluate the performance of CPDP. The performance of the BA method and our approach could be affected by the order of tested modules. Therefore, we randomly changed the order of modules, calculated the AUC 40 times, and computed the average AUC. Following Krishna et al. [7], we set the number of repetitions to 40. Note that when calculating the AUC of the retest (and multiple retests) approach, although the proposed methods updated some non-defective predictions (e.g., t75 and t19), defective predictions were not updated (e.g., t38).

We also used the number of defects found by the prediction (i.e., the number of true positives) to evaluate the performance of each approach. This is because when the number increases, defects that are overlooked during the testing phase can be suppressed but removed later (see Section 3). Note that even if the number of true positives (i.e., found defects) increases, AUC cannot be improved when the number of false negatives also increases. Therefore, we consider both AUC and the number of found defects.

For the evaluation, we defined **_RDIFF_** (relative difference) [6] and **_DIFF_** (difference) as follows:

$$RDIFF(\alpha, \beta) = \frac{criterion\ of\ \beta}{criterion\ of\ \alpha} - 1 \qquad (1)$$

$$DIFF(\alpha, \beta) = criterion\ of\ \beta - criterion\ of\ \alpha \qquad (2)$$

In the equations, the _criterion of α_ denotes the number of found defects by approach α, for instance. For instance, the number of found defects by the baseline approach is 50, and that by the retest one is 55, $RDIFF(baseline, retest)$ is 0.1 (i.e., 10%). Positive values of $DIFF(\alpha, \beta)$ and $RDIFF(\alpha, \beta)$ denote that the approach β improves the performance.

To check the statistical difference in the criteria between the approaches, we applied the Wilcoxon signed-rank test in the analysis.

**Research questions**: To clarify the purpose of the evaluation, we set the following research questions:

- **RQ1**: Is the retest approach more effective than the

---

<div style="text-align:center">**Table 1.** Performance of each approach</div>

(a) 32 projects used as learning data

| Type | AUC | | | | | | Number of found defects | | | | | |
|---|---|---|---|---|---|---|---|---|---|---|---|---|
| | *DIFF* (B, R) | *DIFF* (B, MR) | *DIFF* (R, MR) | *RDIFF* (B, R) | *RDIFF* (B, MR) | *RDIFF* (R, MR) | *DIFF* (B, R) | *DIFF* (B, MR) | *DIFF* (R, MR) | *RDIFF* (B, R) | *RDIFF* (B, MR) | *RDIFF* (R, MR) |
| ε = 0 | **0.014 (0.00)** | **0.016 (0.00)** | 0.002 (0.96) | 2.4% | 2.6% | 0.3% | **1.3 (0.00)** | **1.5 (0.00)** | 0.2 (0.07) | 25.3% | 27.3% | 1.9% |
| ε = 0.1 | **0.023 (0.00)** | **0.030 (0.00)** | 0.007 (0.41) | 3.8% | 5.0% | 1.1% | **3.1 (0.00)** | **5.4 (0.00)** | **2.3 (0.00)** | 46.8% | 103.1% | 56.3% |
| ε = 0.2 | 0.022 (0.10) | 0.023 (0.11) | 0.001 (0.60) | 3.5% | 3.8% | 0.2% | **4.2 (0.00)** | **6.4 (0.00)** | **2.2 (0.00)** | 87.2% | 131.0% | 43.8% |
| ε = 0.3 | 0.011 (0.38) | 0.001 (0.71) | **-0.010 (0.00)** | 1.8% | 0.2% | -1.5% | **3.6 (0.00)** | **4.8 (0.00)** | **1.2 (0.00)** | 35.1% | 47.2% | 12.1% |
| UCB | 0.010 (0.06) | **0.012 (0.04)** | 0.002 (0.91) | 1.7% | 2.1% | 0.3% | **1.4 (0.00)** | **1.8 (0.00)** | **0.4 (0.01)** | 33.1% | 42.2% | 9.1% |
| Average | **0.016 (0.00)** | **0.017 (0.00)** | **0.001 (0.04)** | 2.6% | 2.7% | 0.1% | **2.7 (0.00)** | **4.0 (0.00)** | **1.2 (0.00)** | 45.5% | 70.1% | 24.7% |

(b) 16 projects used as learning data

| Type | AUC | | | | | | Number of found defects | | | | | |
|---|---|---|---|---|---|---|---|---|---|---|---|---|
| | *DIFF* (B, R) | *DIFF* (B, MR) | *DIFF* (R, MR) | *RDIFF* (B, R) | *RDIFF* (B, MR) | *RDIFF* (R, MR) | *DIFF* (B, R) | *DIFF* (B, MR) | *DIFF* (R, MR) | *RDIFF* (B, R) | *RDIFF* (B, MR) | *RDIFF* (R, MR) |
| ε = 0 | 0.007 (0.65) | 0.005 (0.83) | -0.002 (0.17) | 1.1% | 0.8% | -0.3% | **1.0 (0.00)** | **1.1 (0.00)** | 0.1 (0.18) | 13.2% | 14.2% | 1.0% |
| ε = 0.1 | 0.018 (0.15) | 0.013 (0.97) | **-0.004 (0.05)** | 2.8% | 2.1% | -0.7% | **3.0 (0.00)** | **4.0 (0.00)** | **1.0 (0.00)** | 37.8% | 51.1% | 13.2% |
| ε = 0.2 | 0.013 (0.21) | 0.006 (0.80) | **-0.007 (0.02)** | 2.0% | 0.9% | -1.1% | **3.7 (0.00)** | **5.4 (0.00)** | **1.7 (0.00)** | 41.9% | 63.7% | 21.8% |
| ε = 0.3 | 0.008 (0.29) | 0.002 (0.95) | **-0.007 (0.03)** | 1.3% | 0.3% | -1.0% | **3.7 (0.00)** | **4.9 (0.00)** | **1.2 (0.00)** | 35.3% | 47.8% | 12.5% |
| UCB | **0.011 (0.01)** | **0.010 (0.01)** | 0.000 (0.49) | 1.7% | 1.6% | 0.0% | **1.1 (0.00)** | **1.1 (0.00)** | 0.0 (0.32) | 12.4% | 12.6% | 0.2% |
| Average | **0.011 (0.00)** | 0.007 (0.39) | **-0.004 (0.00)** | 1.8% | 1.1% | -0.6% | **2.5 (0.00)** | **3.3 (0.00)** | **0.8 (0.00)** | 28.1% | 37.9% | 9.7% |

(c) 8 projects used as learning data

| Type | AUC | | | | | | Number of found defects | | | | | |
|---|---|---|---|---|---|---|---|---|---|---|---|---|
| | *DIFF* (B, R) | *DIFF* (B, MR) | *DIFF* (R, MR) | *RDIFF* (B, R) | *RDIFF* (B, MR) | *RDIFF* (R, MR) | *DIFF* (B, R) | *DIFF* (B, MR) | *DIFF* (R, MR) | *RDIFF* (B, R) | *RDIFF* (B, MR) | *RDIFF* (R, MR) |
| ε = 0 | 0.002 (0.09) | 0.002 (0.14) | 0.000 (0.10) | 0.4% | 0.3% | 0.0% | **0.4 (0.00)** | **0.4 (0.00)** | 0.0 (1.00) | 2.9% | 2.9% | 0.0% |
| ε = 0.1 | 0.005 (0.34) | **0.021 (0.00)** | **0.016 (0.02)** | 0.7% | 3.3% | 2.5% | **1.0 (0.00)** | **2.6 (0.00)** | **1.7 (0.00)** | 13.5% | 47.0% | 33.5% |
| ε = 0.2 | **0.019 (0.00)** | **0.022 (0.02)** | 0.003 (0.62) | 3.0% | 3.5% | 0.4% | **2.9 (0.00)** | **4.0 (0.00)** | **1.1 (0.00)** | 41.6% | 62.3% | 20.7% |
| ε = 0.3 | 0.012 (0.10) | 0.008 (0.33) | **-0.004 (0.03)** | 1.9% | 1.2% | -0.7% | **2.4 (0.00)** | **2.9 (0.00)** | **0.6 (0.00)** | 24.8% | 30.0% | 5.2% |
| UCB | 0.004 (0.07) | 0.003 (0.15) | **-0.001 (0.03)** | 0.6% | 0.5% | -0.1% | **0.5 (0.00)** | **0.5 (0.00)** | 0.0 (1.00) | 4.1% | 4.1% | 0.0% |
| Average | **0.008 (0.00)** | **0.011 (0.00)** | 0.003 (0.75) | 1.3% | 1.8% | 0.4% | **1.4 (0.00)** | **2.1 (0.00)** | **0.7 (0.00)** | 17.4% | 29.3% | 11.9% |

baseline approach?

- **RQ2**: To what extent is the effect of the multiple retests approach compared to the retest approach?

To answer RQ1, we calculated the differences in AUC between the baseline and retest approaches. Similarly, to answer RQ2, we calculated the differences between the retest and multiple retests approaches.

**Analysis related to RQ1**: Table 1 shows the performance of each approach. As shown in the table, B, R, and MR denote the baseline, retest, and multiple retests approaches. The left side of the table shows the DIFF and RDIFF of the AUC of each approach. Values in parenthesis denote p-values by the Wilcoxon signed-rank test. Light-gray cells mean the p-value is smaller than 0.1, and gray cells with boldface the p-value is smaller than 0.05. Table 2 shows the AUC and the number of defects of each approach. In the table, Proj. means the number of projects used as learning datasets.

All values of *DIFF*(B, R) and *DIFF*(B, MR) of AUC were positive, and the average AUC between the baseline and our approaches was statistically different at 0.05 level, except for when 16 projects were used and the multiple retests approach was applied (see the bottom rows of Table 1). The minimum value of average *RDIFF*(B, R) and *RDIFF*(B, MR) of AUC was 1.1%, and the maximum one was 2.7%. In the work of Kondo et al. [6], the average *RDIFF* of AUC was 1.6% when the best feature reduction technique was applied to defect prediction models such as logistic regression. Compared with the study [6], average *RDIFF* of AUC on our approaches is not very small.

While AUC on each type of BA, such as ε = 0, was not always statistically different. For instance, when 16 projects were used, and the type of BA was ε = 0, AUC was not statistically different between the baseline and our approaches.

The right side of Table 1 shows the *DIFF*(B, R) and *DIFF*(B, MR) of found defects between the baseline and our approaches. The differences were statistically significant at the 0.05 level in all cases, and the average *RDIFF*(B, R) and *RDIFF*(B, MR) of found defects were 17.4% at the minimum (see the bottom rows of Table 1). That is, our approaches significantly improved the number of found defects without degradation of AUC.

Therefore, the retest and the multiple retests approach performed better than the baseline. To answer RQ1, we found that the retest and multiple retests approaches are more effective than the baseline ones.

**Analysis related to RQ2**: On the left side of Table 1, most *DIFF*(R, MR) of AUC were positive when 32 projects were used as learning data. In contrast, many were negative when 8 and 16 projects were used, and the degradations were significantly significant at 0.05 in many cases.

On the right side of Table 1, many of the *DIFF*(R, MR) of found defects were more than zero, and the differences between the retest and multiple retests approaches were statistically significant at 0.05 level in many cases. The average $RDIFF(R, MR)$ in the found defects was 9.7% at the minimum. However, as explained above, AUC was degraded by the multiple retests approach when 8 and 16 projects were used. Hence, applying the multiple retests approach is recommended only when many projects are used as learning data. For RQ2, the effect of the multiple retests approach was smaller than that of the retest approach, especially when many projects are not used as learning data.

**Table 2.** Baseline performance of each approach

| Type | AUC | | | Number of found defects | | |
|---|---|---|---|---|---|---|
| | 32 Proj. | 16 Proj. | 8 Proj. | 32 Proj. | 16 Proj. | 8 Proj. |
| $\varepsilon = 0$ | 0.599 | 0.616 | 0.632 | 8.3 | 9.9 | 10.4 |
| $\varepsilon = 0.1$ | 0.599 | 0.629 | 0.630 | 9.7 | 11.2 | 10.7 |
| $\varepsilon = 0.2$ | 0.612 | 0.628 | 0.627 | 11.1 | 12.2 | 10.8 |
| $\varepsilon = 0.3$ | 0.623 | 0.632 | 0.647 | 12.8 | 13.1 | 12.1 |
| UCB | 0.604 | 0.629 | 0.627 | 9.0 | 10.5 | 10.2 |
| Average | 0.607 | 0.627 | 0.633 | 10.2 | 11.4 | 10.8 |

The results suggested that our approach suppressed defect overlooking without degradation of AUC. However, as explained in Section 3, our approach may increase the retest effort.

## 5. Conclusion

In CPDP, it is challenging to select suitable projects to use for model training. In prior research, Bandit Algorithm (BA) based methods have been used to select projects as learning data. However, in the early stage of software testing, BA can lead to the selection of unsuitable models. The model could predict defective modules as "non-defective", leading to defects overlooking. We proposed a retest based on defect re-prediction to lessen the probability of such overlooking. Our proposed approach is promising because it can have a wide range of applications and is not limited to CPDP.

In the experiment, we evaluated the performance of two types of our approach, the retest approach and the multiple retests approach, compared with the baseline approach, which performs tests only with the ordinal BA method. As evaluation criteria, we used AUC and the number of found defects. As a result, our approach was more effective than the baseline approach. Additionally, the effect of the multiple retests approach was smaller than the retest approach. Although our approach may increase the retesting effort, it is expected to lessen the probability of defect overlooking. In future work, we will apply our approach to other combinations, such as CVDP and online learning methods.

## Acknowledgments

This research is partially supported by the Japan Society for the Promotion of Science [Grants-in-Aid for Scientific Research (C) (No.21K11840, 24K14896, 20H05706).